\documentstyle[12pt]{article}

\begin{document}
\input{psfig.sty}
\begin{flushright}
\baselineskip=12pt
UPR-935-T \\
\end{flushright}

\begin{center}
\vglue 1.5cm
{\Large\bf Spontaneous Physical Compactification
 of the Extra Dimensions}
\vglue 2.0cm
{\Large Tianjun Li~\footnote{E-mail: tli@bokchoy.hep.upenn.edu,
phone: (215) 898-7938, fax: (215) 898-2010.}}
\vglue 1cm
{ Department of Physics and Astronomy \\
University of Pennsylvania, Philadelphia, PA 19104-6396 \\  
U.  S.  A.}
\end{center}

\vglue 1.5cm
\begin{abstract}
We conjecture that the extra dimensions are physical
non-compact at high energy scale or high temperature;
after the symmetry breaking or cosmological phase
transition, the bulk cosmological constant may become negative,
and then, the extra dimensions may become physical compact at low
energy scale. We show this in a
 five-dimensional toy brane model with three parallel 3-branes
and a real bulk scalar whose potential is temperature dependent. 
 We also point out that after the global or gauge symmetry breaking, or
the supersymmetry breaking in supergravity theory, the spontaneous
physical compactification of the extra dimensions might
be realized.

\end{abstract}

\vspace{0.5cm}
\begin{flushleft}
\baselineskip=12pt
April 2000\\
\end{flushleft}
\newpage
\setcounter{page}{1}
\pagestyle{plain}
\baselineskip=14pt

\section{Introduction}

Gauge hierarchy problem is one of the deep problems in
the Standard Model. About three years ago, it was suggested that the large 
compactified extra dimensions may also be the solution to the
gauge hierarchy problem~\cite{AADD}, because a low ($4+n$)-dimensional
Planck scale ($M_X$) may result in the large 4-dimensional Planck scale 
($M_{pl}$) due
to the large physical volume ($V_p^n$)
 of extra dimensions: $M_{pl}^2 = M_X^{2+n} V_p^n$. In addition,
about two years ago, 
 Randall and Sundrum~\cite{LRRS} proposed another scenario
that the extra dimension is an orbifold,
and  the size of extra dimension is not large
but the 4-dimensional mass scale in the Standard Model is
suppressed by an exponential factor from 5-dimensional mass
scale because of the exponential warp factor. Furthermore,  they suggested
that
the fifth dimension might be
coordinate non-compact~\cite{LRRSN}, and there may exist only one
brane with positive tension at origin, but, there  exists the  gauge
hierarchy problem and the physical size of the fifth
dimension is indeed compact. The remarkable aspect of
the second scenario is that it gives rise to a localized graviton field.
Recently, a lot of 5-dimensional models with 3-branes were built [4-5].
We constructed the 
general models with parallel 3-branes on the five-dimensional
space-time, and obtained that the 5-dimensional GUT scale on each brane
can be identified as the 5-dimensional Planck scale, but the
4-dimensional Planck scale is generated from the low 4-dimensional
GUT scale exponentially in our world. Furthermore, 
the models with codimension-1 brane(s) were constructed on 
the six-dimensional and higher dimensional space-time [6-9].  
  
On the other hand, it is well known that five kinds of superstring
theories
 live in the 10-dimensional space-time manifold
and M-theory lives in the 11-dimensional space-time
manifold due to the anomaly cancellation.
In short, it seems that the low energy phenomenology and
 the current fundamental theories favour the existence
of extra dimensions. Therefore, we might
want to ask the question:  why are the extra dimensions physical compact
at low energy scale? From our philosophy, in the early Universe or when
the energy scale or temperature
 is very high, all the space-time dimensions might be phyical 
non-compact, furthermore, all the space dimensions
might be equivalent, i. e., there may be no difference between the
3-dimensional 
space we observe and the extra space dimensions. If this philosophy was
right,
then, we should understand how the extra dimensions become physical
compact at low energy scale.

{\it We conjecture that at high energy scale or high temperature,
 the extra dimensions are physical non-compact. After
the  cosmological phase transition, or the global or gauge symmetry
breaking,
or the supersymmetry breaking in supergravity theory, the bulk
cosmological
constant becomes negative, and then, the extra dimensions become physical
compact.}

In this paper, in order to support our conjecture, we give a
 5-dimensional toy brane model with three paprallel 3-branes and
one real bulk scalar which has temperature dependent potential.
We assume that at high energy scale or temperature,
 the bulk cosmological constant
is zero, because in string theory, it is natural to
take the bulk cosmological constant to be zero since the tree-level 
vacuum energy in the generic critical closed string compactifications
(supersymmetric or not) vanishes, and the zero bulk cosmological
constant is natural in the scenario in which the bulk is supersymmetric
(though the brane need not be), or the quantum corrections to the
bulk are small enough to be neglect in a controlled expansion.
For simplicity,
we also assume that at high energy scale, the metric is flat and diagonal.
When the energy scale or temperature goes down, there is
a second order cosmological phase transition. 
After the cosmological phase transition, the bulk 
cosmological constant becomes negative, and then, the metric has
warp factor, which makes the fifth dimension physical compact. 
In fact, at low energy scale, this model is the same model
discussed previous in~\cite{LTJII}. Of course, this idea can be
generalized to 
all the previous brane models and brane networks with constant bulk
cosmological constant [2-9].

The key point to have the spontaneous physical compactification of
the extra dimensions is that, after
the cosmological phase transition or symmetry breaking, the bulk 
cosmological constant becomes negative,
and then, the extra dimensions become physical compact due to the warp
factor.
Therefore, after the global or gauge symmetry breaking, or
the supersymmetry breaking in supergravity theory, the spontaneous
physical compactification of extra dimensions might
be happened for the cosmological constant might become negative.
And it is interesting to discuss the spontaneous
physical compactification of the extra dimensions in the model
buildings, and in the string theory or M-theory compactifications.

\section{One Toy Model}

In this section, we would like to present a toy brane model
with spontaneous physical compactification of the extra dimension.

We consider the space-time is five-dimensional, and the fifth dimension is
$R^1$. In addition, in order to obtain the cosmological phase transition,
we introduce a real scalar $\phi$ whose bulk potential
is temperature dependent. Assuming that we have three parallel 3-branes
 along the fifth dimension,
and their fifth coordinates are: $-\infty < y_1 < y_2 < y_3 <+\infty$.
We obtain the metric in each brane 
from the five-dimensional metric $g_{AB}$ where  
$A, B = \mu, y$ by restriction
\begin{equation}
\label{smmetric}
g_{\mu \nu}^{(i)} (x^{\mu}) \equiv g_{\mu \nu}(x^{\mu}, y=y_i) ~.~\,
\end{equation}
In this paper, for simplicity, we assume that the metric is diagonal, and 
at high energy scale or temperature, the metric is
$diag(-1, 1, 1, 1, 1)$.

The classical action with a bulk scalar is given by
\begin{eqnarray}
S &=& S_{\rm Bulk} + S_{\rm Brane} 
~,~\,
\end{eqnarray}
\begin{eqnarray}
S_{\rm Bulk} &=&\int d^4 x  ~dy~ \sqrt{-g} \{ 
{1\over 2} M_X^3  R -{1\over 2} \,  \partial_A \phi\,  \partial^A \phi
 - \Lambda (\phi, T)  \} ~,~\,
\end{eqnarray}
\begin{eqnarray}
S_{\rm Brane} &=& \sum_{i=1}^{3} \int d^4 x ~dy~ \sqrt{-g^{(i)}} \{ {\cal
L}_{i} 
-  V_{i}(\phi, T) \} \delta (y-y_i)
~,~\,
\end{eqnarray}
where $M_X$ is the 5-dimensional Planck scale, $T$ is the temperature, 
$\Lambda (\phi, T)$ is the temperature dependent bulk potential for
$\phi$~\cite{MQ}, and 
$V_i(\phi, T)$ where $i=1, 2, 3$ is the brane tension which is dependent
on $\phi$ and $T$.
Our ansatz for $\Lambda(\phi, T)$ and $V_i (\phi, T)$ 
is\footnote{There might exist the terms proportional to $T^5$
and $T^4$ in the $\Lambda(\phi, T)$ and $V_i (\phi, T)$, respectively, 
which will vanish at low temperature. Those additional terms might not
change the compactness of the space, but, will make the discussions of the
spontaneous physical compactification of extra dimension a little bit
complicated. So, as a toy model, we do not consider those terms and 
do not consider how to derive the ansatz Eqs. (5) and (6) here.}
\begin{eqnarray}
\Lambda(\phi, T) &=& D~(T^2-T_o^2)~ \phi^2 +
{{\lambda (T)} \over {4 M_X}} \phi^4 
~,~\,
\end{eqnarray}
\begin{eqnarray}
V_i (\phi, T) &=& C_i (T) ~M_X~ \phi^2 
~,~\,
\end{eqnarray}
where $D$ and  $T_o$ are constant terms, $\lambda (T)$ is a slowly
varying function of $T$. The $C_i(T)$ is also a slowly
varying function and will be determined later.
The temperature dependent potential for $\phi$ has $Z_2$ symmery, which
is invariant under $\phi \leftrightarrow -\phi$.

The solutions to the minimum or maximum of the potential $\Lambda(\phi,
T)$
 are given by
\begin{equation}
<\phi>|_T =0 ~,~\,
\end{equation}
\begin{equation}
<\phi> |_T = \sqrt{{2 D (T_o^2-T^2) M_X }\over\displaystyle \lambda(T)}
~.~\,
\end{equation}
Therefore, the critical temperatue is given by $T_o$. At $T > T_o$, 
the origin $< \phi > = 0$ is the minimum. At the same time, only the 
solution $< \phi > = 0$ does exist. In this case, the bulk cosmological
constant is zero, and all the brane tensions are zero or one can think
that
the brane tensions are very small.
At $T = T_o$, both solutions
collapse at $< \phi > = 0$. At $T < T_o$, the origin $< \phi > = 0$
is the maximum, and the solution
$ <\phi>  = \sqrt{{2 D (T_o^2-T^2) M_X }\over\displaystyle \lambda(T)}$
gives the minimum, so, the $Z_2$ symmetry is
broken. This cosmological phase transition is called of second order
because there is no barrier during the cosmological phase transition.

When $T < T_o$, we can expand the $\phi$ around its minimum, and
write $\phi$ as
\begin{eqnarray}
\phi &=& \phi^{\prime} +
 \sqrt{{2 D (T_o^2-T^2) M_X }\over\displaystyle {\lambda (T)}}
~,~\,
\end{eqnarray}
so, we can express the $\Lambda(\phi, T)$ and $V_i (\phi, T)$
as $\Lambda(\phi^{\prime}, T)$ and $V_i (\phi^{\prime}, T)$,
respectively
\begin{eqnarray}
\Lambda(\phi^{\prime}, T) &=& {{\lambda (T)} \over {4 M_X}}
\left(\phi^{\prime 2} + 2 \phi^{\prime } 
\sqrt{{2 D (T_o^2-T^2) M_X }\over\displaystyle {\lambda (T)}}\right)^2
\nonumber\\&& -
{{D^2 ~M_X}\over {\lambda (T)}} (T_o^2-T^2)^2 ~,~\,
\end{eqnarray}
\begin{eqnarray}
V_i (\phi^{\prime}, T) = C_i (T) M_X 
\left(\phi^{\prime 2} + 2 \phi^{\prime} 
\sqrt{{2 D (T_o^2-T^2) M_X }\over\displaystyle {\lambda (T)}} 
+{{2 D (T_o^2-T^2) M_X }\over\displaystyle {\lambda (T)}}\right)~.~\,
\end{eqnarray}
 
Neglecting the field $\phi^{\prime}$ for
$<\phi^{\prime}> =0$ , during the cosmological
phase transition, we obtain the dominant classical action 
\begin{eqnarray}
S_{\rm Bulk} =\int d^4 x  ~dy~ \sqrt{-g} \left(
{1\over 2} M_X^3  R + {{D^2 ~M_X}\over {\lambda (T)}} (T_o^2-T^2)^2
\right) ~,~\,
\end{eqnarray}
\begin{eqnarray}
S_{\rm Brane} = \sum_{i=1}^{3} \int d^4 x ~dy~ \sqrt{-g^{(i)}} ~\left(
{\cal L}_{i} 
-  {{2~ C_i (T)~ D~ (T_o^2-T^2)~ M_X^2 }\over\displaystyle {\lambda (T)}}
\right)
 \delta (y-y_i) ~.~\,
\end{eqnarray}

Because the calculations are similar to
those in Ref.~\cite{LTJII}, we just give
the result here. 
Assuming that there exists a solution that
 respects  4-dimensional 
Poincare invariance in the $x^{\mu}$-directions, one obtains
the 5-dimensional metric:
\begin{eqnarray} 
ds^2 = e^{- 2 \sigma(y, T)} \eta_{\mu \nu} dx^{\mu} dx^{\nu}
 + dy^2 ~.~\, 
\end{eqnarray}

With this metric, the Einstein equations reduce to
\begin{eqnarray}
\left({{d \sigma} \over {d y}}\right)^2  &=&  
{{D^2~ (T_o^2-T^2)^2 }\over\displaystyle {6 ~M_X^2 ~\lambda (T)}} ~,~\, 
\end{eqnarray} 
\begin{eqnarray}
 {{d^2 \sigma} \over\displaystyle {d^2 y}}  &=&  \sum_{i=1}^{3}
 {{2~C_i (T)~ D~ (T_o^2-T^2) }\over\displaystyle 
{3~ M_X ~\lambda (T)}}
 \delta (y-y_i)
~.~\, 
\end{eqnarray}

The general solution to  above differential equations is
\begin{equation}
\sigma (y, T) = \sum_{i=1}^3 (-1)^{i+1} k (T) |y-y_i| + f[T_o^2-T^2]
~,~\,
\end{equation}
where $f[T_o^2-T^2]$ is the polynomial with variable
 $T_o^2-T^2$ and without
constant term, and $f[T_o^2-T^2]$ is similar to
the constant $c$ in~\cite{LTJII}.

From Eqs. (15-17), we obtain
\begin{eqnarray}
k(T) = \sqrt{1\over {6 \lambda (T)}} 
~{{D (T_o^2-T^2)}\over\displaystyle {M_X}} ~,~ 
\end{eqnarray}
\begin{eqnarray}
C_i (T) = (-1)^{i+1} ~\sqrt{{3 \lambda (T)} \over 2} ~.~\, 
\end{eqnarray}

In addition, for simplicity,
we define that $\sigma (y) \equiv \sigma (y, T=0)$,
$\lambda_o \equiv \lambda (T=0)$,
 $C_i^o \equiv C_i (T=0)$, and $k_o \equiv k (T=0)$. 
After the cosmological phase transition, i. e.,
at very low temperature or energy scale,
we obtain the metric 
\begin{equation}
\sigma (y) = \sum_{i=1}^3 (-1)^{i+1} k_o |y-y_i| + f[T_o^2]
~,~\,
\end{equation}
where
\begin{eqnarray}
k_o = \sqrt{1\over {6 \lambda_o}} ~
{{D T_o^2}\over\displaystyle {M_X}} ~.~\,
\end{eqnarray}
And we have
\begin{eqnarray}
C_i^o = (-1)^{i+1}~ \sqrt{{3 \lambda_o} \over 2} ~.~\, 
\end{eqnarray}

From above metric, we obtain that, after the cosmological phase
transition,
the fifth dimension becomes physical compact
because of the warp factor. And the gauge hierarchy
problem can be solved in this model, as discussed in Ref.~[2, 5].

One might notice that in $\Lambda(\phi^{\prime}, T)$, the lowest order
$\phi^{\prime}$ term is order of $\phi^{\prime 2}$, but in
$V_i (\phi^{\prime}, T)$, the lowest order
$\phi^{\prime}$ term is order of $\phi^{\prime}$. In fact,
  the lowest order
$\phi^{\prime}$ term in $V_i (\phi^{\prime}, T)$ 
 can be order of $\phi^{\prime 2}$, for example, if we defined
the original $V_i (\phi, T)$ as
\begin{eqnarray}
V_i (\phi, T) &=& C_i (T) ~M_X~ \left(\phi^2 - 2 \theta(T_o-T)
\sqrt{{2 D (T_o^2-T^2) M_X }\over\displaystyle {\lambda (T)}} ~\phi
\right)
~,~\,
\end{eqnarray}
or
\begin{eqnarray}
V_i (\phi, T) &=& C_i (T) ~M_X~ \left(\phi^2 - 2 
\sqrt{{2 D |T_o^2-T^2| M_X }\over\displaystyle {\lambda (T)}} ~\phi
\right)
~,~\,
\end{eqnarray}
where $\theta(x)$ is the step function. The definition in Eq. (23)
preserves
the $Z_2$ symmetry before the cosmological phase transition. With
definition
of $V_i (\phi, T)$ in Eq. (23) or Eq. (24),
the discussions of the spontaneous physical compactification of extra
dimension
are similar to the above discussions if one made the
following transformation: $C_i (T) \longrightarrow -C_i (T)$,
and $C_i^o  \longrightarrow -C_i^o$, so, we will not repeat it here.

One can also discuss the first order cosmological phase
transition if one introduced the term $-E(T)~\sqrt {T}~ \phi^3$ in the
bulk potential, however, in this case, there exists barrier during
the cosmological phase transition~\cite{MQ}.

In addition, this idea can be generalized to all the five-dimensional
brane models with parallel 3-branes and the brane networks
discussed in [2-9].

Furthermore, if one considered a compex scalar in the bulk, the
global symmetry will be $U(1)$. Global symmetries are argued to
be broken by quantum gravity effects~\cite{GS}. So, one might need
to consider it as a gauge $U(1)$ symmetry and put additional
particles in the bulk. Similarly, the spontaneous physical
compactification of extra dimension might also be happened in this
scenario.

\section{Discussion and Conclusion} 

The key point in above model to have the spontaneous
 physical compactification of  extra dimension is that, after
the cosmological phase transition or symmetry breaking, the bulk 
cosmological constant becomes negative,
and then, the extra dimension becomes physical compact due to the warp
factor.
 Therefore, in order to search other scenarios where the spontaneous
physical
compactification of extra dimensions may be realized, we need to explore
other possibilities to have negative vacuum energy after the symmetry
breaking.

Obviously, the global or gauge symmetry breaking (Higgs mechanism)
 is one candidate, because after the symmetry breaking, the
cosmological constant will be negative
if the cosmological constant is zero in original theory.
For example, we can put the gauge group and some particles
in the bulk, the gauge symmetry
might be broken by radiative corrections at low energy scale
(similar to the electroweak symmetry breaking in the mSUGRA model), then,
the bulk cosmological constant might become negative and the
spontaneous physical compactification of extra dimensions might be
happened.

The other possibility is the supersymmetry breaking.
As we know, in the supergravity theory, the gravitino mass term
is negative in the potential. And if the supersymmetry was broken,
the gravitino mass will give a negative contribution to the cosmological
constant. So, the cosmological constant might be negative after
the supersymmetry breaking, and then, we might have the 
spontaneous physical compactification of extra dimensions .
Of course,  at high energy scale, the supersymmetry is preserved, so, the
cosmological constant is zero and the extra dimensions are physical
non-compact. 

 How to realize the spontaneous physical compactification of extra
dimensions
due to the global or gauge symmetry breaking, or the supersymmetry
breaking
 in the model buildings  and in the string theory or
M-theory compactifications is an interesting subject and deserves
further study.

In short, we conjecture that at high energy scale or high temperature,
 the extra dimensions are physical non-compact. After
the  cosmological phase transition, or the global or gauge symmetry
breaking,
or the supersymmetry breaking in supergravity theory, the bulk
cosmological
constant becomes negative, and then, the extra dimensions become physical
compact. We show this in a
 five-dimensional toy brane model with three parallel 3-branes
and a real scalar whose bulk potential is temperature dependent.
In the mean time, the gauge hierarchy problem can be solved in the toy
model. 

\section*{Acknowledgments}
This research was supported in part by the U.S.~Department of Energy under
 Grant No.~DOE-EY-76-02-3071.

\end{document}